\documentclass[aps,prl,twocolumn,preprintnumbers,superscriptaddress]{revtex4}
\usepackage{graphicx}% Include figure files
\usepackage{dcolumn}% Align table columns on decimal point
\usepackage{epstopdf}
\begin{document}

\preprint{Ver. 3}

\title{Comment on ``Quantum Criticality and Nodal Superconductivity in the FeAs-Based Superconductor KFe$_2$As$_2$''}


\author{Taichi Terashima}
\author{Motoi Kimata}
\author{Nobuyuki Kurita}
\affiliation{National Institute for Materials Science, Tsukuba, Ibaraki 305-0003, Japan}
\affiliation{JST, Transformative Research-Project on Iron Pnictides (TRIP), Chiyoda, Tokyo 102-0075, Japan}
\author{Hidetaka Satsukawa}
\author{Atsushi Harada}
\author{Kaori Hazama}
\affiliation{National Institute for Materials Science, Tsukuba, Ibaraki 305-0003, Japan}
\author{Motoharu Imai}
\affiliation{National Institute for Materials Science, Tsukuba, Ibaraki 305-0003, Japan}
\affiliation{JST, Transformative Research-Project on Iron Pnictides (TRIP), Chiyoda, Tokyo 102-0075, Japan}
\author{Akira Sato}
\affiliation{National Institute for Materials Science, Tsukuba, Ibaraki 305-0003, Japan}
\author{Kunihiro Kihou}
\author{Chul-Ho Lee}
\author{Hijiri Kito}
\author{Hiroshi Eisaki}
\author{Akira Iyo}
\affiliation{JST, Transformative Research-Project on Iron Pnictides (TRIP), Chiyoda, Tokyo 102-0075, Japan}
\affiliation{National Institute of Advanced Industrial Science and Technology (AIST), Tsukuba, Ibaraki 305-8568, Japan}
\author{Taku Saito}
\affiliation{Department of Physics, Chiba University, Chiba 263-8522, Japan}
\author{Hideto Fukazawa}
\author{Yoh Kohori}
\affiliation{JST, Transformative Research-Project on Iron Pnictides (TRIP), Chiyoda, Tokyo 102-0075, Japan}
\affiliation{Department of Physics, Chiba University, Chiba 263-8522, Japan}
\author{Hisatomo Harima}
\affiliation{JST, Transformative Research-Project on Iron Pnictides (TRIP), Chiyoda, Tokyo 102-0075, Japan}
\affiliation{Department of Physics, Graduate School of Science, Kobe University, Kobe, Hyogo 657-8501, Japan}
\author{Shinya Uji}
\affiliation{National Institute for Materials Science, Tsukuba, Ibaraki 305-0003, Japan}
\affiliation{JST, Transformative Research-Project on Iron Pnictides (TRIP), Chiyoda, Tokyo 102-0075, Japan}


\date{\today}

%\begin{abstract}
%\end{abstract}

%\pacs{71.18.+y, 71.27.+a, 74.70.Tx}
%\keywords{}

\maketitle

%\section{}
%\subsection{}
%\subsubsection{}

% If in two-column mode, this environment will change to single-column
% format so that long equations can be displayed. Use
%\begin{widetext}
%\end{widetext}

\newcommand{\ud}{\mathrm{d}}
\def\degree{\kern-.2em\r{}\kern-.3em}

In a recent Letter \cite{Dong10PRL}, Dong \textit{et  al}. have observed a $T^{1.5}$ dependence of resistivity $\rho$ in KFe$_2$As$_2$ at the upper critical field $B_{c2}$ = 5 T parallel to the $c$ axis and have suggested the existence of a field-induced quantum critical point (QCP) at $B_{c2}$.  The value of $B_{c2}$ = 5 T was determined from the onset of a resistive transition.  It is much higher than a perviously reported value of $B_{c2}$ = 1.25 T \cite{Terashima09JPSJKFA}, which was determined from a midpoint.  The origin of the large difference in $B_{c2}$ can be attributed to broad transitions (see the inset of Fig. 1 of Ref.~\onlinecite{Dong10PRL}, the $\rho(T)$ curve at zero field shows the onset at 4.8 K and zero resistivity at 2.5 K), which indicate distribution of the transition temperature and $B_{c2}$ due to inhomogeneity in the sample.  Since the observed exponent 1.5 is some kind of average over the inhomogeneous sample, it is questionable to relate it to a QCP.  To clarify this, let us assume that some portion of a sample becomes quantum critical at $B_{c2}$ = 5 T and that $\rho$ in that portion varies as $T^{1.5}$.  Since $B_{c2}$ for the rest of the sample is below 5 T, the rest is in the Fermi liquid regime and exhibits a $T^2$ dependence of $\rho$.  Then, one can not expect $\rho$ measured across the sample varies as $T^{1.5}$. 

\begin{figure}
\includegraphics[width=7cm]{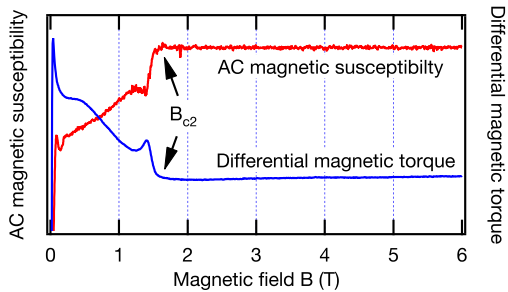}
\caption{\label{fig1}(color online).  AC magnetic susceptibility at $T <$ 0.1 K and field derivative of magnetic torque at $T$ = 0.04 K for nearly $B \parallel c$.  $B_{c2}$ is approximately marked.}   
\end{figure}

\begin{figure}
\includegraphics[width=9cm]{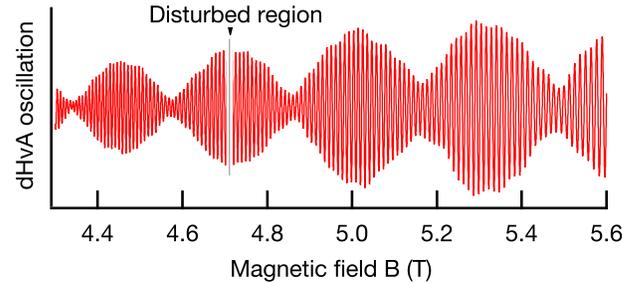}
\caption{\label{fig1}(color online).  dHvA oscillations for $B \parallel c$ at $T$ = 0.04 K.  The magnet power supply exhibits a small output current jump at $B$ = 4.72 T, which disturbs dHvA oscillations as indicated by the vertical grey line.}   
\end{figure}

%Recently we have observed de Haas-van Alphen (dHvA) oscillations in KFe$_2$As$_2$ using high-quality single crystals with resistivity ratios of up to $\sim$600 \cite{Terashima10condmat}.    In the following, we substantiate the argument against the proposed QCP at $B_{c2}$ = 5 T, mentioning data obtained for these dHvA-quality crystals.

We substantiate below the argument against the proposed QCP, mentioning data obtained for high-quality single crystals with resistivity ratios of up to $\sim$600, which exhibit de Haas-van Alphen (dHvA) oscillations \cite{Terashima10condmat}.

Figure 1 shows magnetic susceptibility and torque measured near $B \parallel c$.  It indicates that $B_{c2}$ for $B \parallel c$ is roughly 1.6 T.  At the field of the claimed $B_{c2}$ of 5 T, the bulk of KFe$_2$As$_2$ is situated far away from $B_{c2}$.

We show dHvA oscillations of KFe$_2$As$_2$ in a field range near the proposed QCP ($B$ = 5 T) in Fig. 2.  dHvA oscillations are clearly observed from above the proposed QCP to below it with no anomaly.  The existence of dHvA oscillations is usually regarded as a sign of a Fermi liquid.  (Since Dong \textit{et al}. show a $T^{1.5}$ dependence at $B$ = 3 T for a new sample in the Reply \cite{Dong_reply}, we show data down to 3 T in Ref.~\onlinecite{EPAPS}) In Ref.~\onlinecite{Terashima10condmat}, we have examined the field dependence of effective masses $m^*$'s for quasiparticles on two sheets of the Fermi surface; $m^*$'s are constant in the investigated field range $7 < B < 17.65$ T within the errors of $\pm$6--10\% for one sheet and $\pm$3\% for the other.  This contrasts sharply with the two-fold increase in the coefficient $A$ of the $T^2$ term of $\rho$ from 14.5 to 6 T toward the proposed QCP reported in Ref. 1.  The latter corresponds to $\sim$40\% increase in $m^*$, assuming that $A \sim (m^*)^2$ as usual.

In summary, the observation of a $T^{1.5}$ dependence of $\rho$ in the sample showing broad resistive transitions does not constitute evidence for a QCP, and recent dHvA results do not support the proposed QCP.

\begin{figure*}
{\Large Supplemental material} 
\includegraphics[width=17cm]{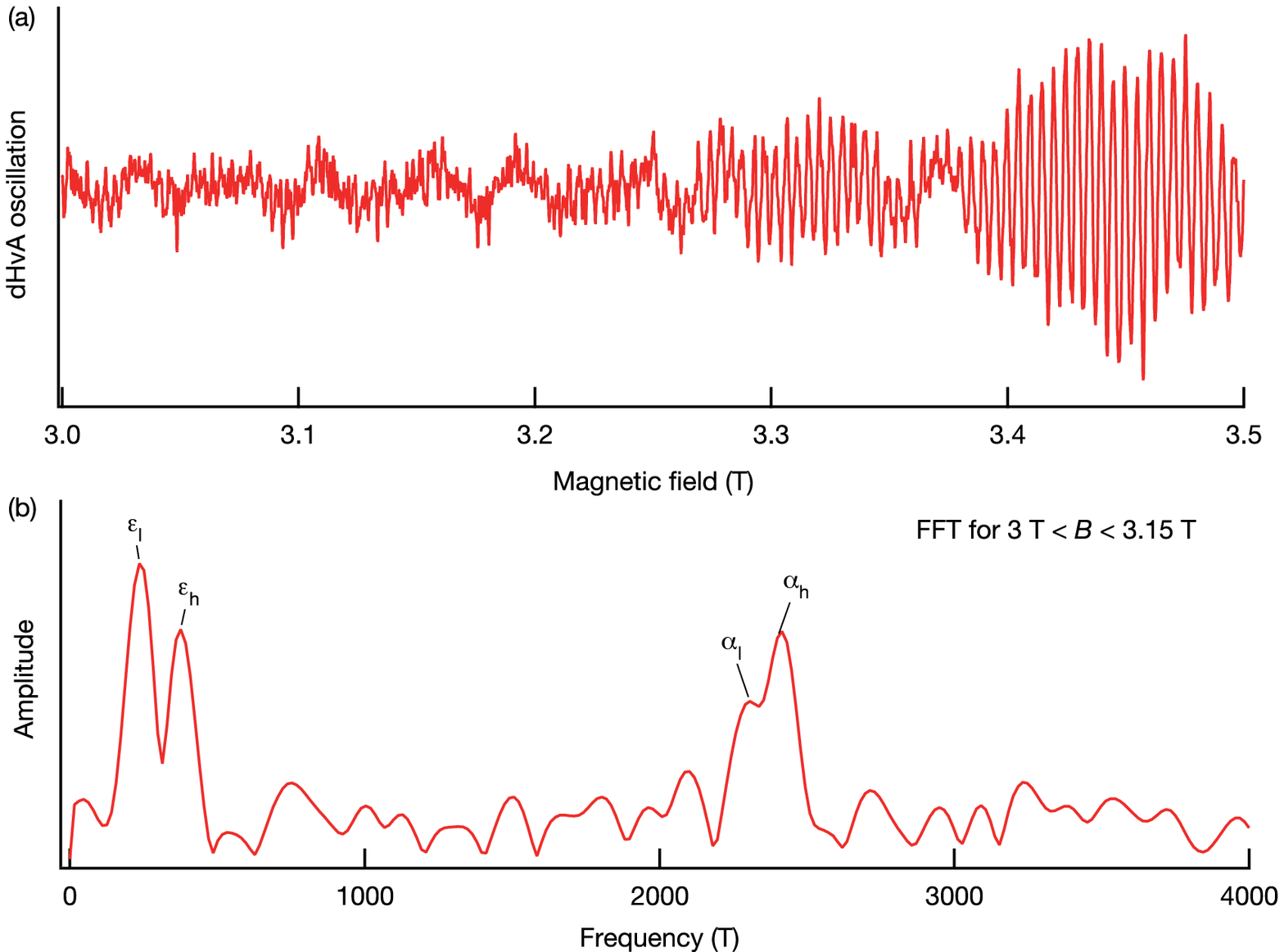} \newline
Fig.  (a) dHvA oscillation in KFe$_2$As$_2$ at $T$ = 0.04 K for $B \parallel c$ down to $B$ = 3 T.  (b) Fourier transform (in 1/$B$) for a field range 3 T $< B <$ 3.15 T (not 3.5 T).  Four dHvA frequencies, $\epsilon_l$, $\epsilon_h$, $\alpha_l$, and $\alpha_h$, are clearly resolved even for this very low field region.
\end{figure*}

\end{document}